\documentstyle[12pt]{article}
\input epsf
\newcommand{\postscript}[2] 
{\setlength{\epsfxsize}{#2\hsize}\centerline{\epsfbox{#1}}}
\newcommand{\nc}{\newcommand}
\nc{\bg}{B. Grz${{{\rm a}_{}}_{}}_{\hskip -0.18cm\varsigma}$dkowski}
\nc{\lsp}{\;\;\;\;\;\;\;\;}
\nc{\non}{\nonumber}
\nc{\barx}{\bar{x}}
\nc{\pbarn}{\;\hbox {pb}}
\nc{\hc}{\hbox {h.c.}}
\nc{\re}{\hbox {Re}}
\nc{\mev}{\hbox {MeV}} \nc{\gev}{\;\hbox {GeV}}
\def\gesim{\lower0.5ex\hbox{$\:\buildrel >\over\sim\:$}} 
\def\lesim{\lower0.5ex\hbox{$\:\buildrel <\over\sim\:$}} 
\nc{\prd}[3]{{\it Phys.\ Rev.}\ {{\bf D{#1}} (#2), #3}}
\nc{\prl}[3]{{\it Phys.\ Rev.\ Lett.}\ {{\bf {#1}} (#2), #3}}
\nc{\plb}[3]{{\it Phys.\ Lett.}\ {{\bf B{#1}} (#2), #3}}
\nc{\npb}[3]{{\it Nucl.\ Phys.}\ {{\bf B{#1}} (#2), #3}}
\nc{\ptp}[3]{{\it Prog.\ Theor.\ Phys.}\ {{\bf {#1}} (#2), #3}}
\nc{\zfp}[3]{{\it Z.\ Phys.}\ {{\bf C{#1}} (#2), #3}}
\nc{\mpla}[3]{{\it Mod.\ Phys.\ Lett.}\ {{\bf A{#1}} (#2), #3}}
\nc{\rmp}[3]{{\it Rev.\ Mod.\ Phys.}\ {{\bf {#1}} (#2), #3}}
\nc{\ijmpa}[3]{{\it Int.\ J.\ of\ Mod.\ Phys.}\
               {{\bf A{#1}} (#2), #3}}
\nc{\ttbar}{t\bar{t}}         \nc{\bbbar}{b\bar{b}}
\nc{\tanb}{\tan \beta}
\nc{\twbdec}{t\to W^+ b}
\nc{\tbwbdec}{\bar{t}\to W^- \bar{b}}
\nc{\epem}{e^+e^-}            \nc{\eett}{\epem \to \ttbar}
\nc{\sigeett}{\sigma_{e\bar{e}\to\ttbar}}
\nc{\wpwm}{W^+W^-}            
\nc{\tbar}{\bar{t}}           \nc{\bbar}{\bar{b}}
\nc{\wpp}{W^+}
\nc{\mt}{m_t} \nc{\mts}{m_t^2}\nc{\mw}{m_W} \nc{\mws}{m_W^2}
\nc{\mz}{m_Z} \nc{\mzs}{m_Z^2}
\nc{\ttbardec}{\ttbar \to W^+W^-\bbbar}
\nc{\wwbb}{W^+W^-\bbbar}
\nc{\sm}{SM}
\nc{\cw}{\cos\theta_W}        \nc{\sw}{\sin\theta_W}
\nc{\sws}{\sin^2\theta_W}
\nc{\sig}{\sigma_{tot}}
\nc{\lp}{\ell^+}              \nc{\lm}{\ell^-}
\nc{\fp}{F_+}                 \nc{\fm}{F_-}
\nc{\gp}{G_+}                 \nc{\gm}{G_-}
\nc{\fpm}{F_{\pm}}            \nc{\gpm}{G_{\pm}}
\nc{\epsl}{\epsilon_L}

\begin{document}
\pagestyle{empty} \setlength{\footskip}{2.0cm}
\setlength{\oddsidemargin}{0.5cm} \setlength{\evensidemargin}{0.5cm}
\renewcommand{\thepage}{-- \arabic{page} --}
\def\mib#1{\mbox{\boldmath $#1$}}
\def\bra#1{\langle #1 |}      \def\ket#1{|#1\rangle}
\def\vev#1{\langle #1\rangle} \def\dps{\displaystyle}
   \def\thebibliography#1{\centerline{REFERENCES}
     \list{[\arabic{enumi}]}{\settowidth\labelwidth{[#1]}\leftmargin
     \labelwidth\advance\leftmargin\labelsep\usecounter{enumi}}
     \def\newblock{\hskip .11em plus .33em minus -.07em}\sloppy
     \clubpenalty4000\widowpenalty4000\sfcode`\.=1000\relax}\let
     \endthebibliography=\endlist
   \def\sec#1{\addtocounter{section}{1}\section*{\hspace*{-0.72cm}
     \normalsize\bf\arabic{section}.$\;$#1}\vspace*{-0.3cm}}
\vspace*{-1.6cm}\noindent
\hspace*{10.8cm}IFT-07-96\\
\hspace*{10.8cm}TOKUSHIMA 96-01\\
\hspace*{10.8cm}(hep-ph/9604301)\\

\vspace*{.5cm}

\begin{center}
{\large\bf Energy Spectrum of Secondary Leptons in $\mib{e}^+
\mib{e}^-\!\to \mib{t}\bar{\mib{t}}$}

\vskip 0.2cm
{\large\it --- Non-Standard Interactions and CP violation ---}
\end{center}

\vspace*{1cm}
\begin{center}
\renewcommand{\thefootnote}{*)}
{\sc Bohdan
GRZ${{{\rm A}_{}}_{}}_{\hskip -0.18cm\varsigma}
$DKOWSKI$^{\: a),\: }$}\footnote{E-mail address:
\tt bohdan.grzadkowski@fuw.edu.pl}
and
\renewcommand{\thefootnote}{**)}
{\sc Zenr\=o HIOKI$^{\: b),\: }$}\footnote{E-mail address:
\tt hioki@ias.tokushima-u.ac.jp}
\end{center}

\vspace*{1.2cm}
\centerline{\sl $a)$ Institute for Theoretical Physics,\ Warsaw 
University}
\centerline{\sl Ho\.za 69, PL-00-681 Warsaw, POLAND} 

\vskip 0.3cm
\centerline{\sl $b)$ Institute of Theoretical Physics,\ 
University of Tokushima}
\centerline{\sl Tokushima 770, JAPAN}

\vspace*{1.5cm}
\centerline{ABSTRACT}

\vspace*{0.4cm}
\baselineskip=20pt plus 0.1pt minus 0.1pt
The process of top-quark pair production at future high-energy $e^+
e^-$ linear colliders has been investigated as a possible test of
physics beyond the Standard Model. Non-standard interactions have
been assumed both for the production and for the subsequent decay of
the top quarks. The energy spectrum of the single lepton $\ell^\pm$
and the energy correlation of $\lp$ and $\lm$ emerging from the
process $e^+e^-\!\to t\bar{t}\to \ell^\pm X/\ell^+\ell^-X$ are
calculated. The energy-spectrum asymmetry of  $\lp$ and $\lm$ is
considered as a measure of $C\!P$ violation. An optimal method to
determine whether $C\!P$ violation occurs in the production or in the
decay processes is proposed.

\vfill
\newpage
\renewcommand{\thefootnote}{\sharp\arabic{footnote}}
\pagestyle{plain} \setcounter{footnote}{0}
\baselineskip=21.0pt plus 0.2pt minus 0.1pt

\sec{Introduction}
$C\!P$ violation is a challenging significant problem in electroweak
physics. Decays of $D$ and $B$ mesons have been extensively
investigated for this purpose. In the near future we are expecting
much more fruitful experimental data from $B$-factories under
construction. On the other hand, the top-quark production may be
another efficient source of information on $C\!P$ violation once
Large Hadron Collider (LHC) and/or Next Linear Collider (NLC) are
constructed, as discussed in $[1\: -\:6]$.

It is relevant to notice that the amount of $C\!P$ violation provided
by the Cabibbo-Kobayashi-Maskawa mechanism for the top-quark sector
is tiny, therefore $C\!P$ violation in this sector offers a wide
window to look for physics beyond the Standard Model (SM).
Furthermore, the top quark is expected to give us a unique
opportunity to study quark interactions much more directly thanks to
its extremely large mass, $m_t^{exp}=180 \pm 12$ GeV \cite{top}.
Since the top quark is so heavy it decays as a single quark before
forming bound states, therefore it is possible to avoid complicated
non-perturbative effects brought through fragmentation processes in a
case of lighter quarks.

Since $\ttbar$ pairs are produced through the vector-boson exchange,
the handedness of $t$ and $\bar{t}$ must be the same. Consequently,
the helicities of $\ttbar$ would be $(+-)$ or $(-+)$ if the top mass
were much smaller than $\sqrt{s}$. However, since the observed $m_t$
is far from being negligible at any accelerators in the planning
stage, we will also face copious production of $(++)$ and $(--)$
states. For example, $\sigma_{tot}(e^+e^-\!\to\ttbar)$ is estimated
to be 0.60 pb for $\sqrt{s}=$500 GeV (and $m_t=$180 GeV) within the
SM, in which $N(-+):N(+-):N(--):N(++)$ is $4.8:3.4:0.9:0.9$, where
$N(\cdots)$ denotes the number of $\ttbar$ pairs with the indicated
helicities (cf. this ratio would be about $6.1:3.9:O(10^{-5}):
O(10^{-5})$ if $m_t$ were the same as $m_b$).

We can use this fact to explore $C\!P$ properties of the $\ttbar$
state: $\ket{-+}$ and $\ket{+-}$ are $C\!P$ self-conjugate while
$\ket{--}$ and $\ket{++}$ transform into each other under $C\!P$
operation as
$$
\hat{C}\!\hat{P}\ket{\mp\mp}=\hat{C}\ket{\pm\pm}=\ket{\pm\pm}.
$$
This indicates that the difference between $N(--)$ and $N(++)$ could
be a useful measure of $C\!P$ violation $[4\: -\:6]$, although what
we can observe in experiments are not the top quarks but products of
their subsequent decays. Fortunately we know that the semileptonic
decays can serve as an efficient top-quark-spin analyzers
\cite{topspin}. Indeed, the energy spectrum of $\ell^+$ and $\ell^-$
in $e^+e^-\!\to t\bar{t} \to b\bar{b}W^+W^-\to b\bar{b}\ell^+\nu
\ell^- \bar{\nu}$ can be a good measure of $N(--)-N(++)$, as we will
see. One can understand it qualitatively since:
\begin{itemize}
\item[(1)] The large top mass requires a predominantly longitudinal
$W$ in $t\to bW$ since $\bar{b}\gamma_{\mu}(\gamma_5)t\cdot
\varepsilon^{\mu}\sim m_t\bar{b}(\gamma_5)t$ when $\varepsilon^{\mu}=
\varepsilon_L^{\mu}\sim k^{\mu}$ ($\varepsilon$ and $k$ are the
polarization and the four-momentum of $W$, respectively).
\item[(2)] The produced $b$ ($\bar{b}$) is left-handed (right-handed)
in the SM since $m_b/\sqrt{s}\ll$1.
\item[(3)] Because of (1) and (2), $W^+$'s three-momentum prefers to
be parallel (anti-parallel) to that of $t(+)$($t(-)$), where
$t(\cdots)$ expresses a top with the indicated helicity. Consequently
$\ell^+$ in the $t(+)$ decay becomes more energetic than in the
$t(-)$ decay, while it is just opposite for the $\bar{t}$ decay,
i.e., $\bar{t}(-)$ produces more energetic $\ell^-$ than $\bar{t}(+)$
does.
\item[(4)] Therefore, we expect larger number of energetic $\ell^+$
($\ell^-$) for $N(--)<N(++)$ (for $N(--)>N(++)$).
\end{itemize}

The leptonic energy spectrum has been studied in the existing 
literature \cite{CKP,AS}. However, in those articles,
$C\!P$-violating interactions were assumed only in the $t\bar{t}
\gamma/Z$ vertices, and the standard-model vertex was used for the
$t\to bW$ decay. In this paper, in order to perform a consistent
analysis we compute the spectrum assuming that both the $t\bar{t}
\gamma/Z$ vertices and the $tbW$ vertex include non-standard
$C\!P$-violating form factors. Concerning the $W$ decays, we shall
treat them as in the SM since it is known through various
charged-current processes that the $W$ couplings with light fermions
are successfully described within the SM. That is, we shall assume
here that {\it only the top-quark interactions may be modified by
physics beyond the SM.}

The paper is organized as follows. In sec.\ 2 we will describe a
formalism for the energy spectrum calculation together with some
related SM results. In sec.\ 3 we will consider the top-quark decay
with non-standard interactions. Section\ 4 will contain a derivation
of the lepton-energy spectrum with $C\!P$ violation present both in
the production and in the decay. Then, in sec.\ 5, we will discuss
how to measure $C\!P$ violation effectively and propose an optimal
method to disentangle effects originating from the production and
from the decay. In the Appendix, explicit forms of some functions
used in the text will be presented.
  
\sec{The lepton-energy spectrum}
Before proceeding to actual study of $C\!P$ violation, let us briefly
describe the formalism which we use in this paper, and show the
related standard-model calculations.

We will treat all the fermions except the top-quark as massless and
adopt a technique developed by Kawasaki, Shirafuji and Tsai
\cite{technique}. This is a useful method to calculate distributions
of final particles appearing in a process of production and
subsequent decay. This technique is applicable when the narrow-width
approximation
$$
\left|\,{1\over{p^2-m^2+im{\mit\Gamma}}}\,\right|^2
\simeq{\pi\over{m{\mit\Gamma}}}\delta(p^2 -m^2)
$$
can be adopted for the decaying intermediate particles. In fact, this
is very well satisfied for the production and subsequent decays of
$t$ and $W$ since ${\mit\Gamma}_t\simeq$ 175$\;\mev (\mt/\mw)^3\ll
\mt$ and ${\mit\Gamma}_W=2.08\pm 0.07$ GeV \cite{PDG} $\ll M_W$.

Adopting this method, one can derive the following formula for the
inclusive distribution of the single-lepton $\ell^+$ in the reaction
$\eett$\ \cite{AS}:
\begin{eqnarray}
&&\frac{d^3\sigma}{d^3 p_\ell/(2p_\ell^0)}(\epem \to \ell^+ + \cdots)
\non \\
&&\ \ \ \ \ \ \ \ \ \ \ \ \ \ \ \ \ 
=4\int d{\mit\Omega}_t
\frac{d\sigma}{d{\mit\Omega}_t}(n,0)\frac{1}{{\mit\Gamma}_t}
\frac{d^3{\mit\Gamma}_\ell}{d^3 p_\ell/(2p_\ell^0)}(t\to b\ell^+\nu),
\label{master}
\end{eqnarray}
where ${\mit\Gamma}_\ell$ is the leptonic width of {\it unpolarized}
top and $d\sigma(n,0)/d{\mit\Omega}_t$ is obtained from the angular
distribution of $\ttbar$ with spins $s_+$ and $s_-$ in $\eett$,
$d\sigma(s_+,s_-)/d{\mit\Omega}_t$, by the following replacement:
\begin{equation}
s^\mu_+ \to n^\mu=\left(g^{\mu \nu}-\frac{p_t^\mu p_t^\nu}{\mts}
\right)\frac{\mt}{p_t p_\ell}p_{\ell\,\nu}
\lsp{\rm and}\lsp s_- \to 0.
\label{replacement}
\end{equation}
(Exchanging the roles of $s_+$ and $s_-$ and reversing the sign of
$n^\mu$, we get the distribution of $\ell^-$.)

Following ref. \cite{AS}, let us introduce the rescaled
lepton-energy, $x$, by
\begin{equation}
x\equiv
\frac{2 E_\ell}{\mt}\left(\frac{1-\beta}{1+\beta}\right)^{1/2},
\label{def-x}
\end{equation}
where $E_\ell$ is the energy of $\ell$ in $\epem$ c.m. frame and
$\beta=\sqrt{1-4\mts/s}$ ($s\equiv(p_{e^+}+p_{e^-})^2)$. We also
define three parameters $D_V$, $D_A$ and $D_{V\!A}$ as
\begin{eqnarray*}
&&D_V=|v_e v_t d-\frac23|^2 +|a_e v_t d|^2, \\
&&D_A=|v_e a_t d|^2 +|a_e a_t d|^2,         \\
&&D_{V\!A}=v_e a_t d(v_e v_t d-\frac23)^* +a_e a_t d(a_e v_t d)^*,
\end{eqnarray*}
by using the standard-model neutral-current parameters of $e$ and
$t$: $v_e=-1+4\sin^2\theta_W$, $a_e=-1$, $v_t=1-(8/3)\sin^2\theta_W$,
and $a_t=1$, and a $Z$-propagator factor
$$
d\equiv\frac{s}{s-M_Z^2+iM_Z{\mit\Gamma}_Z}
\frac{1}{16\sin^2\theta_W\cos^2\theta_W}.
$$

Then, the $x$ spectrum is given in terms of these quantities by
\begin{eqnarray}
\frac{1}{B_\ell\sigeett}{\frac{d\sigma}{dx}}^{\!\pm}\equiv
\frac{1}{B_\ell\sigeett}
\frac{d\sigma}{dx}(\epem \to \ell^\pm+\cdots)=f(x)+\eta\: g(x).
\end{eqnarray}
Here $\sigeett\equiv\sigma_{tot}(e^+e^-\!\to\ttbar)$, $B_\ell$ is the
branching ratio for $t\to \ell +\cdots$ ($\simeq 0.22$ for $\ell=e,
\mu$), $f(x)$ and $g(x)$ are functions derived in \cite{AS}, which we
give in the Appendix, and $\eta$ is defined as
$$
\eta\equiv
{{4\:{\rm Re}(D_{V\!A})}\over{(3-\beta^2)D_V +2\beta^2 D_A}}.
$$
\newpage
\begin{figure}[h]
\postscript{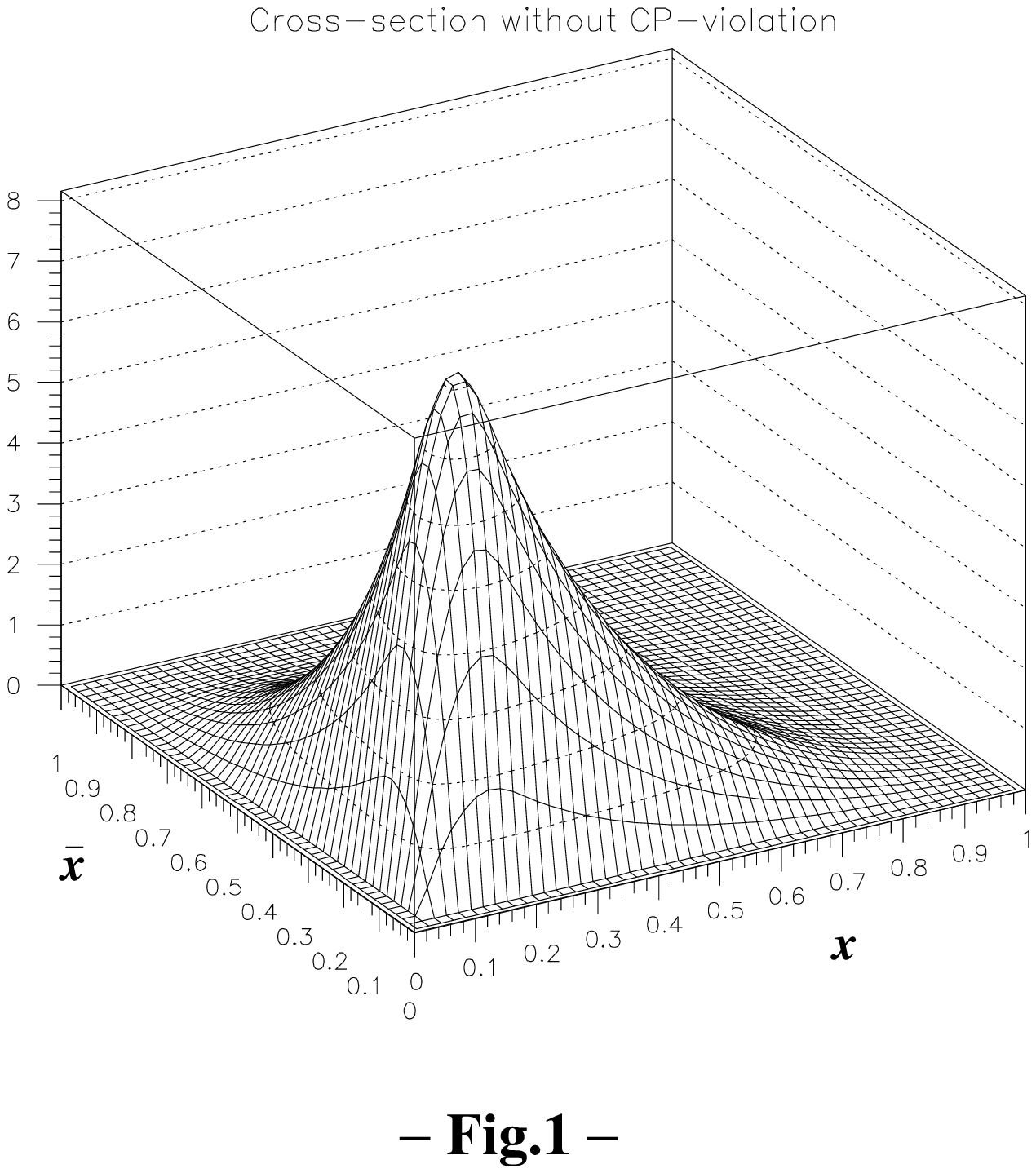}{0.75}
\caption{The normalized $(x,\;\bar{x})$ distribution $\sigma^{-1}\:
d^2\sigma/(dx \:d\bar{x})$ without $C\!P$ violation.}
\end{figure}
\noindent
$f(x)$ and $g(x)$ satisfy the following normalization conditions:
\begin{equation}
\int f(x)dx=1 \lsp {\rm and} \lsp \int g(x)dx=0.
\end{equation}

Applying the same technique, we get the following energy correlation
of $\lp$ and $\lm$:
\begin{equation}
\frac{1}{B_\ell^2 \sigeett} \frac{d^2\sigma}{dx\;d\barx}=
S_0 (x,\barx),
\end{equation}
where $x$ and $\barx$ are the rescaled energies of $\ell^+$ and
$\ell^-$ respectively, and
$$
S_0 (x,\barx)=f(x)f(\barx)+\eta'g(x)g(\barx)
+\eta [\,f(x)g(\barx)+g(x)f(\barx)\,]
$$
with $\eta'$ being defined as
$$
\eta'\equiv\frac{1}{\beta^2}\frac{(1+\beta^2)D_V+2\beta^2 D_A}
{(3-\beta^2)D_V+2\beta^2 D_A}.
$$
Clearly, the  $(x,\bar{x})$ distribution is symmetric in $x$ and
$\barx$, which is a sign of $C\!P$ symmetry. The distribution is
presented in fig.1 for $\sqrt{s}=500$ GeV and the SM parameters
$\sin^2\theta_W=0.2325$, $M_W=80.26$ GeV, $M_Z=91.1884$ GeV,
${\mit\Gamma}_Z=2.4963$ GeV and $m_t=180$ GeV.
\setlength{\footskip}{2cm}

\sec{Non-standard interactions and the top-quark decay}
We will assume that all non-standard effects in the production
process can be represented by the photon and $Z$-boson exchange in
the $s$-channel in the following way: \footnote{Two other possible
    form factors do not contribute in the limit of zero electron 
    mass.} 
\begin{equation}
{\mit\Gamma}^\mu_v=\frac{g}{2}\bar{u}(p_t)
\biggl[\,\gamma^\mu(A_v-B_v\gamma_5)
+\frac{(p_t-p_{\bar{t}})^\mu}{2\mt}(C_v-D_v\gamma_5)\,\biggr]v(p_t),
\label{vtt}
\end{equation}
where $v=\gamma$ or $Z$ and $g$ is the SU(2) gauge-coupling constant.
A non-zero value of $D_v$ is a signal of $C\!P$ violation.

For the on-shell $W$, we will adopt the following parameterization of
the $tbW$ vertex suitable for the $\twbdec$ and $\tbwbdec$ decays:
\begin{eqnarray}
&&{\mit\Gamma}^{\mu}=-{g\over\sqrt{2}}V_{tb}\:
\bar{u}(p_b)\biggl[\,\gamma^{\mu}(f_1^L P_L +f_1^R P_R)
-{{i\sigma^{\mu\nu}k_{\nu}}\over M_W}
(f_2^L P_L +f_2^R P_R)\,\biggr]u(p_t),\ \ \ \ \\
&&\bar{\mit\Gamma}^{\mu}=-{g\over\sqrt{2}}V_{tb}^*\:
\bar{v}(p_t)\biggl[\,\gamma^{\mu}(\bar{f}_1^L P_L +\bar{f}_1^R P_R)
-{{i\sigma^{\mu\nu}k_{\nu}}\over M_W}
(\bar{f}_2^L P_L +\bar{f}_2^R P_R)\,\biggr]v(p_b),
\end{eqnarray}
where $P_{L/R}=(1\mp\gamma_5)/2$, $V_{tb}$ is the $(tb)$ element of
the Kobayashi-Maskawa matrix and $k$ is the momentum of $W$. Again,
because $W$ is on shell, the two additional form factors do not
contribute. One can show that \cite{cprelation} 
\begin{equation}
f_1^{L,R}=\pm\bar{f}_1^{L,R},\lsp f_2^{L,R}=\pm\bar{f}_2^{R,L},
\label{cprel}
\end{equation}
where upper (lower) signs are those for $C\!P$-conserving
(-violating) contributions. Therefore any $C\!P$-violating observable
defined for the top-quark decay must be proportional to $f_1^{L,R}-
\bar{f}_1^{L,R}$ or $f_2^{L,R}-\bar{f}_2^{R,L}$.

We shall consider here the top-quark decay with the above
non-standard-interaction terms. Assuming that
$\stackrel{\scriptscriptstyle(-)}{f_1^L}-1$,
$\stackrel{\scriptscriptstyle(-)}{f_1^R}$,
$\stackrel{\scriptscriptstyle(-)}{f_2^L}$ and
$\stackrel{\scriptscriptstyle(-)}{f_2^R}$ are small and keeping only
linear terms, we obtain for the double differential spectrum in $x$
and $\omega\equiv(p_t-p_\ell)^2/m_t^2$ the following result:
\begin{equation}
\frac{1}{{\mit\Gamma}_t}
\frac{d^2{\mit\Gamma}_\ell}{dxd\omega}(t\to b\ell^+ \nu)
=\frac{1+\beta}{\beta}\;\frac{3 B_\ell}{W}
\omega\left[1+2{\rm Re}(f_2^R)\sqrt{r}\left(\frac{1}{1-\omega}-
\frac{3}{1+2r} \right)\right], \label{t-decay}
\end{equation}
where
$$
W\equiv(1-r)^2(1+2r),\lsp r\equiv M_W^2/m_t^2.
$$
An analogous formula for $\bar{t}\to\bar{b}\ell^-\bar{\nu}$ holds
with $f_2^R$ replaced by $\bar{f}_2^L$.

\sec{$\mib{C}\!\mib{P}$ violation in the production and in the decay 
processes}
Combining the results of the previous sections, we obtain the
lepton-energy spectrum for $\epem \to l^\pm+\cdots$ with the
non-standard $C\!P$-violating terms as \footnote{Since our main
    interest is in $C\!P$ violation, we dropped all the
    $C\!P$-conserving non-standard terms in eq.(\ref{vtt}).} 
\begin{eqnarray}
\frac{1}{B_\ell\sigeett}{\frac{d\sigma}{dx}}^{\!\pm}
=\fpm(x)+(\eta\mp\xi)\gpm(x), \label{s-dis}
\end{eqnarray}
where 
\begin{eqnarray*}
&&\xi \equiv\frac{1}{(3-\beta^2)D_V+2\beta^2 D_A} \\
&&\ \ \ \ \ \times
\frac{-1}{\sin\theta_W}{\rm Re}\biggl[\:\frac23\: D_\gamma
+\frac{s^2}{(s-M_Z^2)^2+M_Z^2{\mit\Gamma}_Z^2}
\frac{(v_e^2+a_e^2)v_t}{64\sin^3\theta_W\cos^3\theta_W}D_Z
\\
&&\ \ \ \ \
-\frac{s(s-M_Z^2)}{(s-M_Z^2)^2+M_Z^2{\mit\Gamma}_Z^2}
\biggl(\,\frac{v_e v_t}{16\sin^2\theta_W\cos^2\theta_W}D_\gamma
+\frac{v_e}{6\sin\theta_W\cos\theta_W}D_Z \,\biggr)\:\biggr],
\end{eqnarray*}
which characterizes the $C\!P$ violation in the $t\bar{t}$ production
process,\footnote{This point will become much clearer in later
   discussions (see eq.(\ref{delta})).}\ 
and $\fpm(x)$ and $\gpm(x)$ are defined as
\begin{eqnarray}
&&\fp(x)=f(x)+{\rm Re}(f_2^R)\:\delta\! f(x), \ \
  \gp(x)=g(x)+{\rm Re}(f_2^R)\:\delta g(x), \\
&&\fm(x)=f(x)+{\rm Re}(\bar{f}_2^L)\:\delta\! f(x), \ \
  \gm(x)=g(x)+{\rm Re}(\bar{f}_2^L)\:\delta g(x), 
\end{eqnarray}
with $\delta\!f(x)$ and $\delta g(x)$ being given in the Appendix.
Note that $\fpm(x)$ and $\gpm(x)$ satisfy the same normalization
conditions as $f(x)$ and $g(x)$:
\begin{equation}
\int \fpm(x)dx=1 \lsp {\rm and} \lsp \int \gpm(x)dx=0.
\end{equation}
%
\begin{figure}[h]
\postscript{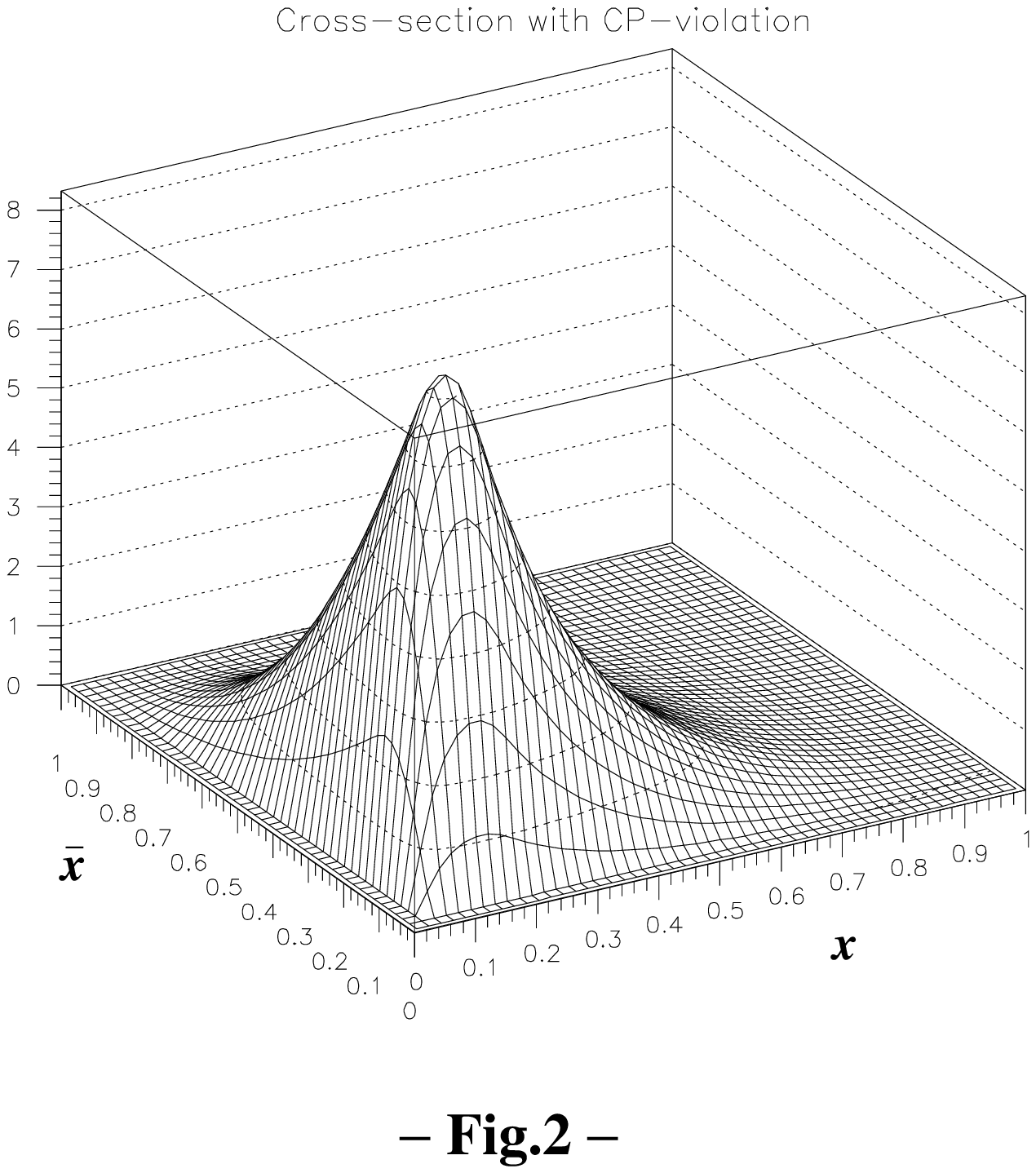}{0.75}
\caption{The normalized $(x,\;\bar{x})$ distribution $\sigma^{-1}\:
d^2\sigma/(dx \:d\bar{x})$ with $C\!P$ violation for ${\rm Re}
(D_{\gamma})={\rm Re}(D_Z)={\rm Re}(f_2^R)=-{\rm Re}(\bar{f}_2^L)=
0.2$.}
\end{figure}

The $(x,\barx)$ distribution receives extra pieces, which are
anti-symmetric in $x$ and $\barx$:
\begin{equation}
\frac{1}{B_\ell^2 \sigeett} \frac{d^2\sigma}{dx\;d\barx}
=S(x,\barx)+\xi A_{\xi}(x,\barx), \label{CPV-dis}
\end{equation}
where $S(x,\barx)$ is obtained through replacement of $f(x)$ and
$g(x)$ ($f(\barx)$ and $g(\barx)$) by $\fp(x)$ and $\gp(x)$
($\fm(\barx)$ and $\gm(\barx)$) in $S_0 (x,\barx)$ as
\begin{eqnarray}
&&S(x,\barx)=\fp(x)\fm(\barx)+\eta'\gp(x)\gm(\barx) \non \\
&&\ \ \ \ \ \ \ \ \ \ \ \
+\eta [\,\fp(x)\gm(\barx)+\gp(x)\fm(\barx)\,] \non 
\end{eqnarray}
and
\begin{eqnarray}
A_{\xi}(x,\barx)=\fp(x)\gm(\barx)-\gp(x)\fm(\barx). \non 
\end{eqnarray}

The distribution is shown for ${\rm Re}(D_{\gamma})={\rm Re}(D_Z)=
{\rm Re}(f_2^R)=-{\rm Re}(\bar{f}_2^L)=0.2$ in fig.2, where all the
SM parameters are the same as in fig.1. Since we assumed that the
non-standard interactions are not strong, the two distributions in
figs.1 and 2
\begin{figure}[h]
\postscript{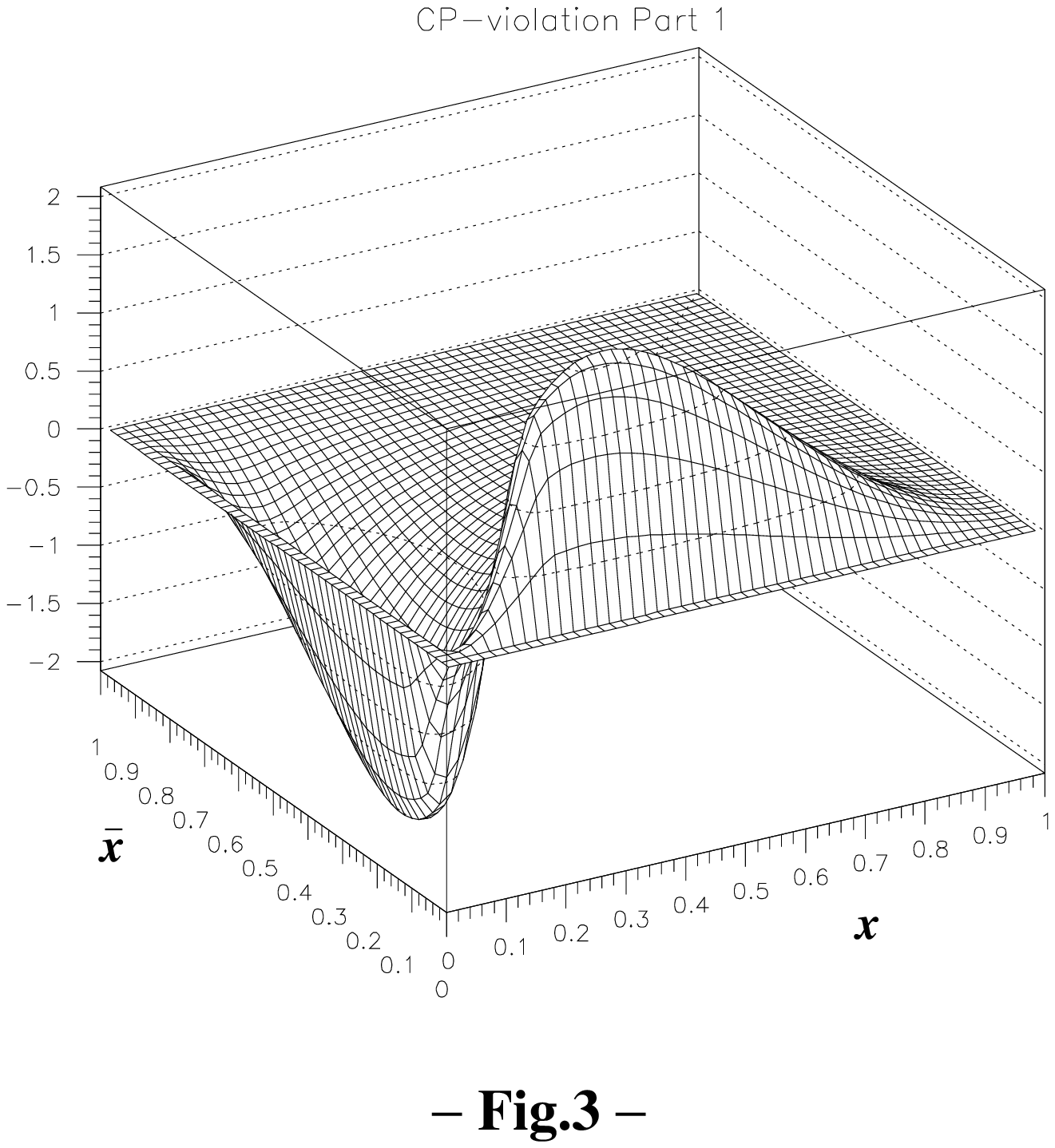}{0.75}
\caption{The $C\!P$-violating function $A_{\xi}(x,\bar{x})$ given in
eq.(18).}
\end{figure}
look similar to each other at first sight. However, looking
carefully at the contour lines, we find that the distribution in
fig.2 is not symmetric in $x$ and $\bar{x}$, which is a sign of
$C\!P$ violation. In order to show more explicitly the both
$C\!P$-violating contributions, we re-express the right-hand side of
eq.(\ref{CPV-dis}) as
\begin{equation}
S_0(x,\barx)+\xi A_\xi(x,\barx)
+\frac12 {\rm Re}(f_2^R-\bar{f}_2^L) A_f(x,\barx), \label{CPV-part}
\end{equation}
and show $A_{\xi,\,f}(x, \bar{x})$ in figs.3 and 4 respectively.
\begin{figure}[h]
\postscript{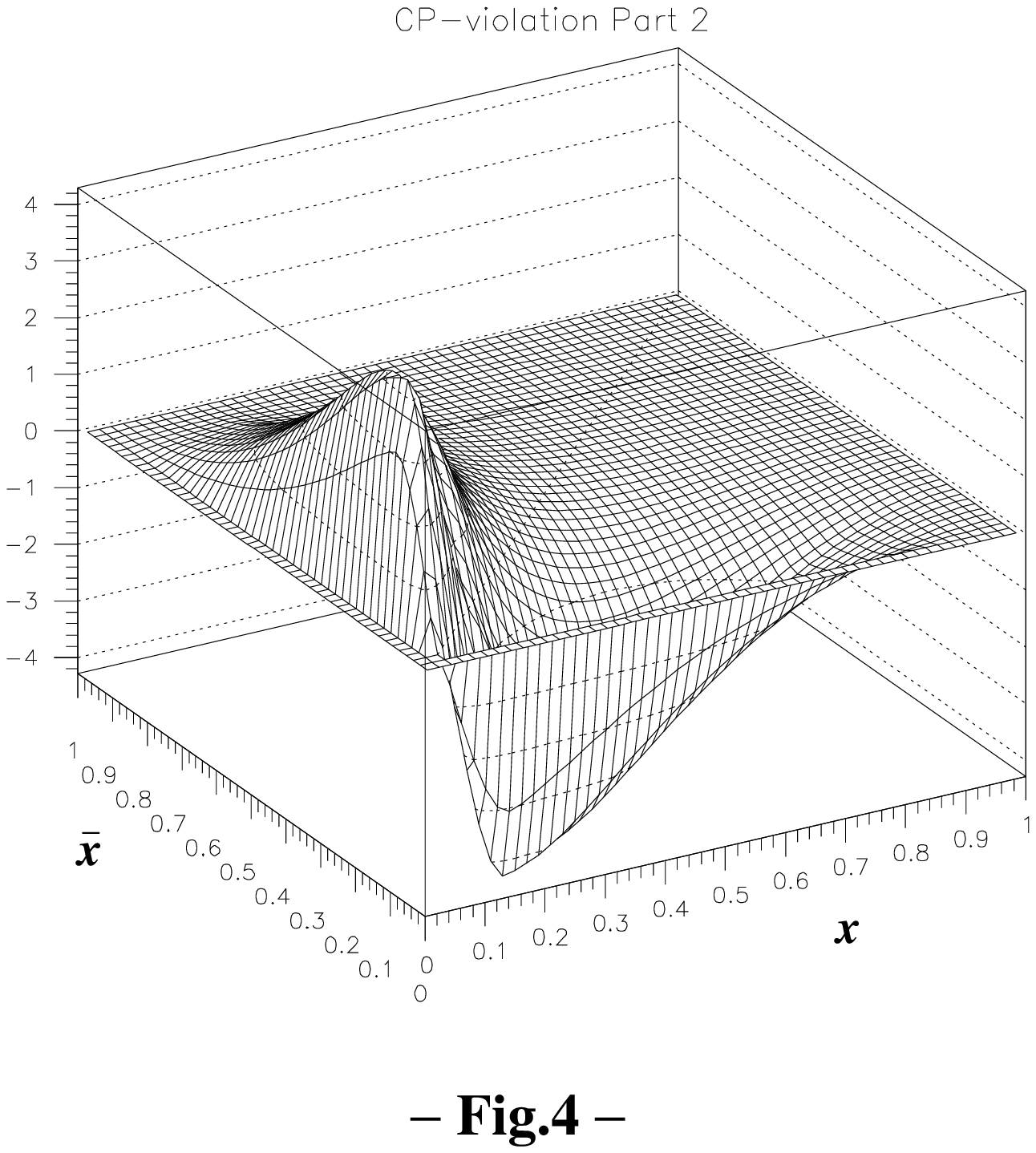}{0.75}
\caption{The $C\!P$-violating function $A_f(x,\bar{x})$ given in
eq.(19).}
\end{figure}

Both $A_\xi(x,\bar{x})$ and $A_f(x,\barx)$ are anti-symmetric in $x$
and $\barx$. It is worth to notice that within the approximation
adopted in this paper (keeping linear terms in the non-standard
couplings)
\begin{eqnarray}
&&A_\xi(x,\barx)= f(x)g(\bar{x})-g(x)f(\bar{x}), \\
&&A_f(x,\barx)= \delta\! f(x)f(\bar{x})-f(x)\delta\! f(\bar{x})
+\eta'\:[\:\delta g(x)g(\bar{x})-g(x)\delta g(\bar{x})\:] \non \\
&&\ \ \ \ \ \ \ \ \ \ \
+\eta\:[\:\delta\! f(x)g(\bar{x})-f(x)\delta g(\bar{x})
+\delta g(x)f(\bar{x})-g(x)\delta\! f(\bar{x})\:].
\end{eqnarray}

\sec{Measurements of $\mib{C}\!\mib{P}$ violation}
As mentioned in the Introduction,
\begin{equation}
\delta\equiv\frac{N(--)-N(++)}{N(all)}
\label{helasym}
\end{equation}
is a measure of $C\!P$ violation in the production process. One can
show, assuming the dominance of $\gamma$ and $Z$ exchange in the
$s$-channel, that $\delta$ is related to the parameter $\xi$
introduced in eq.(\ref{s-dis}):
\begin{equation}
\delta=-\beta \xi. \label{delta}
\end{equation}

If there was no $C\!P$ violation in the $tbW$ vertex, the
energy-spectrum asymmetry $a(x)$ defined as
\begin{equation}
a(x)\equiv\frac{d\sigma^-/dx-d\sigma^+/dx}{d\sigma^-/dx+d\sigma^+/dx}
\label{asy}
\end{equation}
would be given by a simple form
$$
a(x)=-\frac{\delta}{\beta} \frac{g(x)}{f(x)+\eta\: g(x)},
$$
and may serve as a useful observable to measure $C\!P$ violation. 
However, when the $C\!P$-violating contributions to the $tbW$ vertex
are taken into account, it becomes
$$
a(x)=\frac{-2(\delta/\beta)\,g(x) +\re{(f_2^R-\bar{f}_2^L)}
[\,\delta\! f(x)+\eta\: \delta g(x)\,]}{2\,[\,f(x)+\eta\: g(x)\,]}.
$$
Therefore, it turns out that $a(x)$ is not a direct measure of the
helicity asymmetry $\delta$. Measuring a differential asymmetry is a
challenging task since $a(x)$ has not been integrated over the energy
and therefore the expected statistics cannot be high.

We shall find more appropriate observables to measure $C\!P$
violation in the production (expressed by $\xi$) and that in the
decay process (expressed by $\re{(f_2^R-\bar{f}_2^L)}$) individually.
It shall be useful to write down explicit expressions for the single
lepton spectrum:
\begin{eqnarray}
&&\frac{1}{\sigma^+}{\frac{d\sigma}{dx}}^{\!+}=f(x)+\eta\,g(x)
-\xi\,g(x) +\re{(f_2^R)}[\,\delta\! f(x)+\eta\: \delta g(x)\,],\\
&&\frac{1}{\sigma^-}{\frac{d\sigma}{dx}}^{\!-}=f(x)+\eta\,g(x)
+\xi\,g(x) +\re{(\bar{f}_2^L)}[\,\delta\! f(x)+\eta\: \delta g(x)\,],
\end{eqnarray}
where 
$$\sigma^\pm\equiv
\int dx\:{\frac{d\sigma}{dx}}^{\!\pm}=B_\ell \: \sigeett.
$$
Now, following the methods developed in ref.\ \cite{optimalization}
one can show that in order to maximize statistical significance of
the $C\!P$-violating signal the following observables should be
applied:
\begin{equation}
{\cal{O}}_{ttV}^\pm=\frac{1}{\sigma^\pm}
\int dx\: {\frac{d\sigma}{dx}}^{\!\pm}{\mit\Theta}_{ttV}(x),\ \ \
{\cal{O}}_{tbW}^\pm=\frac{1}{\sigma^\pm}
\int dx\: {\frac{d\sigma}{dx}}^{\!\pm}{\mit\Theta}_{tbW}(x),
\label{observ}
\end{equation}
where we shall use
\begin{equation}
{\mit\Theta}_{ttV}(x)=\frac{g(x)}{f(x)+\eta\: g(x)},\ \ \
{\mit\Theta}_{tbW}(x)=
\frac{\delta\! f(x) + \eta\: \delta g(x)}{f(x)+\eta\: g(x)},
\end{equation}
as the weighting functions. ${\cal{O}}_{ttV}^\pm$ and
${\cal{O}}_{tbW}^\pm$ are the most sensitive observables for a
measurement of $\xi$ and $\re{(f_2^R-\bar{f}_2^L)}$ respectively.
Once those ${\cal{O}}_{ttV,\,tbW}^\pm$ are experimentally determined,
we are able to obtain $\xi$ and $\re{(f_2^R-\bar{f}_2^L)}$ through
\begin{eqnarray}
2\xi
&=&\frac{c({\cal{O}}_{ttV}^+ -{\cal{O}}_{ttV}^-)
        -a({\cal{O}}_{tbW}^+ -{\cal{O}}_{tbW}^-)}{a^2-bc} \non\\
&=&\int dx\:\left[\,\frac{1}{\sigma^+}{\frac{d\sigma}{dx}}^{\!+}
               -\frac{1}{\sigma^-}{\frac{d\sigma}{dx}}^{\!-}\,\right]
{\mit\Omega}_\xi (x),
\label{cpobser1}
\end{eqnarray}
\begin{eqnarray}
\re{(f_2^R-\bar{f}_2^L)}
&=&\frac{a({\cal{O}}_{ttV}^+ -{\cal{O}}_{ttV}^-)
        -b({\cal{O}}_{tbW}^+ -{\cal{O}}_{tbW}^-)}{a^2-bc} \non\\
&=&\int dx\:\left[\,\frac{1}{\sigma^+}{\frac{d\sigma}{dx}}^{\!+}
               -\frac{1}{\sigma^-}{\frac{d\sigma}{dx}}^{\!-}\,\right]
{\mit\Omega}_f (x),\label{cpobser2}
\end{eqnarray}
for 
$$
{\mit\Omega}_\xi(x)=
\frac{c\:{\mit\Theta}_{ttV}(x)-a\:{\mit\Theta}_{tbW}(x)}{a^2-bc},
\ \ \
{\mit\Omega}_f (x)=\frac{a\:{\mit\Theta}_{ttV}(x)
-b\:{\mit\Theta}_{tbW}(x)}{a^2-bc},
$$
where 
\begin{eqnarray}
a&\equiv&
\int dx\: \frac{g(x)[\:\delta\! f(x) + \eta\: \delta g(x)\:]}{f(x)
+ \eta\: g(x)},\non\\ 
b&\equiv&
\int dx\: \frac{g^2(x)}{f(x) + \eta\: g(x)}, \non\\ 
c&\equiv&
\int dx\: \frac{[\:\delta\! f(x) + \eta\: \delta g(x)\:]^2}{f(x)
+ \eta\: g(x)}.\non
\end{eqnarray}

Using eqs.(\ref{cpobser1},\ \ref{cpobser2}) one can calculate the
statistical errors for $2\xi$ and $\re{(f_2^R-\bar{f}_2^L)}$
measurements:
\begin{eqnarray}
{\mit\Delta}_i=\left[{\mit\Delta}_{i+}^2+{\mit\Delta}_{i-}^2
\right]^{1/2}\ \ (i=\xi,\: f),
\end{eqnarray}
where ${\mit\Delta}_{i\pm}$ denotes the statistical error for
$(\sigma^\pm)^{-1}{\dps\int}dx\:{\mit\Omega}_i(x)(d\sigma^\pm/dx)$
measurement given by
\begin{equation}
{\mit\Delta}_{i\pm}=\frac{1}{\sqrt{N_\ell}}\left[\:
\frac{1}{\sigma^\pm}\int dx\:
{\mit\Omega}_{i}^2(x) {\frac{d\sigma}{dx}}^{\!\pm}-
\left(\frac{1}{\sigma^\pm}\int dx\: {\mit\Omega}_i(x)
{\frac{d\sigma}{dx}}^{\!\pm}\right)^2\:\right]^{1/2}
\end{equation}
with $N_\ell$ being the total number of events with  one lepton
$\ell^\pm$ for the integrated luminosity $L$. Therefore the
statistical significances $N_{S\!D}^{ttV}$ and $N_{S\!D}^{tbW}$ with
which the presence of a nonzero value of $\xi$ and
$\re{(f_2^R-\bar{f}_2^L)}$ may be ascertained are
\begin{eqnarray}
N_{S\!D}^{ttV}=|\,2\xi\:|/{\mit\Delta}_\xi
\ \ \ \ {\rm and}\ \ \ \
N_{S\!D}^{tbW}=|\,\re{(f_2^R-\bar{f}_2^L)}\:|/{\mit\Delta}_f.
\label{NSD}
\end{eqnarray}

Within the approximation adopted in this paper, we may use the
standard-model formula to estimate the size of ${\mit\Delta}$'s,
consequently we have ${\mit\Delta}_{i+}={\mit\Delta}_{i-}$.
Eventually we obtain for the errors:
\begin{eqnarray}
{\mit\Delta}_{\xi}=17.44/\sqrt{N_\ell}
\ \ \ \ {\rm and}\ \ \ \ 
{\mit\Delta}_f=10.06/\sqrt{N_\ell}.
\end{eqnarray}
There are two quantities relevant for the experimental potential of
NLC, namely the total integrated luminosity $L$ and the tagging
efficiency for an observation of $\ttbar$ pairs $\epsilon_{tt}$ in
various decay channels. Since they enter the statistical significance
in a combination $\sqrt{\epsilon_{tt} L}$, it will be useful to adopt
a notation $\epsl\equiv\sqrt{\epsilon_{tt} L}$ and parameterize our 
results in terms of $\epsl$. Table 1 shows $\epsilon_{tt}$ (in \%)
necessary to achieve a desired $\epsl$ corresponding to a given
luminosity $L$ (note that $\epsilon_{tt}\leq B_\ell\simeq$ 22 \% for
the single-lepton-inclusive final state).
\setlength{\footskip}{2cm}

\def\ss{\scriptsize}
\def\sst{\scriptstyle}
\def\ssst{\scriptscriptstyle}
\renewcommand{\arraystretch}{1.2}
\begin{table}
\vspace*{-0.4cm}
\begin{center}
\begin{tabular}{@{\vrule width0.8pt}c@{\vrule width0.8pt}r|r|r|r|r
@{\vrule width0.8pt}}\noalign{\hrule height0.8pt}
\ $\epsl$ {\ss (pb}$^{\ssst{-1/2}}${\ss )}~
&\multicolumn{5}{c@{\vrule width0.8pt}}
{$L$ {\ss (10}$^{\ssst{4}}${\ss pb}$^{\ssst{-1}}${\ss )}}\\
\cline{2-6}
   &   2.   &  5.   & 10.   & 15.   & 20.~  \\
                                         \noalign{\hrule height0.8pt}
 50&~12.5\ &\ 5.0\ &\ 2.5\ &\ 1.7\ &\ 1.3~ \\ \hline
100&  $-$  &  20.0 & 10.0  &  6.7  &  5.0~ \\ \hline
200&  $-$  &  $-$  &  $-$  &  $-$  & 20.0~ \\ 
                                         \noalign{\hrule height0.8pt}
\end{tabular}\\ \vspace*{0.3cm}
{\bf -- Table 1 --}
\end{center}
\vspace*{-0.5cm}
\caption{$\epsilon_{tt}$ (in \%) necessary to achieve a desired
$\epsl$ corresponding to a given luminosity $L$ for the
single-lepton-inclusive final state.}
\end{table}

Some rough estimations of the efficiency are available in the
literature \cite{efficiency}. For instance, $\epsilon_{tt}=15\:\%$
may be obtained for 4 jets + one charged lepton. If $L=40\ 
{\rm fb}^{-1}$ is achieved,\footnote{$L=10-100\ {\rm fb}^{-1}$ is
    used in, e.g., \cite{JLC}.}\ 
we will obtain $\epsl=$ 77.5 $\pbarn^{-1/2}$. Since $\sigeett=0.60
\pbarn $ for $\mt=180 \gev$, we have $\sqrt{N_\ell}(=
\sqrt{\epsilon_{tt}L\sigeett})$= 0.77$\:\epsl/\pbarn^{-1/2}$. 
Therefore we can compute the minimal values for $|\,\xi\:|$ and 
$|\,\re{(f_2^R-\bar{f}_2^L)}\:|$ observable at a desired statistical
significance for a given $\epsl$ as
\begin{eqnarray}
&&|\,\xi\:|^{min}=11.3(N_{S\!D}^{ttV}/\epsilon_L\pbarn^{1/2})
\ {\rm and}\ 
|\,\re{(f_2^R-\bar{f}_2^L)}\:|^{min}=13.1(N_{S\!D}^{tbW}/\epsilon_L
\pbarn^{1/2}). \non\\ 
&& \label{NSD2}
\end{eqnarray}
Having $\epsl=77.5 \pbarn^{-1/2}$ we will be able to test $\xi$ down
to 0.44 and $\re{(f_2^R-\bar{f}_2^L)}$ down to 0.51 at three standard
deviations. Some other typical values are given in tables 2 and 3. 
\setlength{\footskip}{2cm}

\begin{table}[h]
\vspace*{-0.4cm}
\begin{center}
\begin{tabular}{@{\vrule width0.8pt}c@{\vrule width0.8pt}c|c|c|c|c
@{\vrule width0.8pt}}\noalign{\hrule height0.8pt}
\ $\epsl$ {\ss (pb}$^{\ssst{-1/2}}${\ss )}~
&\multicolumn{5}{c@{\vrule width0.8pt}}{$N_{S\!D}^{ttV}$}
\\ \cline{2-6}
&~~1&2&3&4&5~~\\ \noalign{\hrule height0.8pt}
 50 &~ 0.23 & 0.45 & 0.68 & 0.90 & 1.13~ \\ \hline
100 &~ 0.11 & 0.23 & 0.34 & 0.45 & 0.57~ \\ \hline
200 &~ 0.06 & 0.11 & 0.17 & 0.23 & 0.28~ \\ 
                                         \noalign{\hrule height0.8pt}
\end{tabular}\\ \vspace*{0.3cm}
{\bf -- Table 2 --}
\end{center}
\vspace*{-0.5cm}
\caption{Minimal values of $|\,\xi\:|$ observable at a desired
statistical significance $N_{S\!D}^{ttV}$ for a given $\epsl$.}
\end{table}

\begin{table}
\vspace*{-0.4cm}
\begin{center}
\begin{tabular}{@{\vrule width0.8pt}c@{\vrule width0.8pt}c|c|c|c|c
@{\vrule width0.8pt}}\noalign{\hrule height0.8pt}
\ $\epsl$ {\ss (pb}$^{\ssst{-1/2}}${\ss )}~ 
&\multicolumn{5}{c@{\vrule width0.8pt}}{$N_{S\!D}^{tbW}$}
\\ \cline{2-6}
&~~1&2&3&4&5~~\\ \noalign{\hrule height0.8pt}
 50 &~ 0.26 & 0.52 & 0.79 & 1.05 & 1.31~ \\ \hline
100 &~ 0.13 & 0.26 & 0.39 & 0.52 & 0.66~ \\ \hline
200 &~ 0.07 & 0.13 & 0.20 & 0.26 & 0.33~ \\ 
                                         \noalign{\hrule height0.8pt}
\end{tabular}\\ \vspace*{0.3cm}
{\bf -- Table 3 --}
\end{center}
\vspace*{-0.5cm}
\caption{Minimal values of $|\,\re{(f_2^R-\bar{f}_2^L)}\:|$
observable at a desired statistical significance $N_{S\!D}^{tbW}$ for
a given $\epsl$.}
\end{table}

We have not considered any background yet. However, since majority of
the single-lepton-inclusive final states is made of 4 jets + one
charged lepton + missing energy, the background seems to be easy
under control. In this case the final state could be fully
reconstructed since there is only one neutrino, 3 jets must add up to
a priori known top-quark mass, and two of them must have the $M_W$
invariant mass. Because of those constraints we would assume that the
background could be neglected. 

{\it Within the SM non-zero $\xi$ and $\re{(f_2^R-\bar{f}_2^L)}$ may
appear at the two-loop level.} Therefore, {\it an observation of
non-zero $\xi$ or $\re{(f_2^R-\bar{f}_2^L)}$ would be a strong
indication for non-standard physics.}

\sec{Summary}
Next-generation linear colliders of $e^+ e^-$ will provide a cleanest
environment for studying top-quark interactions. There, we shall be
able to perform detailed tests of the top-quark couplings to the
vector bosons and either confirm the SM simple generation-repetition
pattern or discover some non-standard interactions.

In this paper, we have studied the non-standard $C\!P$-violating
interactions in the $t\bar{t}$ productions and their subsequent
decays. $C\!P$ violation has been parameterized by $\xi$
(eqs.(\ref{s-dis}, \ref{helasym}, \ref{delta})) and 
$\re{(f_2^R-\bar{f}_2^L)}$ (eq.(\ref{cprel})) for the production and
decay process, respectively. If the top-quark decay was described by
the SM interactions (as it was done in the previous study
\cite{CKP,SP,AS}), then we would have a compact useful formula for a
measurement of $C\!P$ violation in the $t\bar{t}\gamma/Z$ vertices
via the final-lepton energy-asymmetry (\ref{asy}). However, in
general, $C\!P$ violation may also enter through the top-decay
process at the same strength as it does for the production.
Therefore, we have assumed the most general $C\!P$-violating
interactions both in the production and in the decay vertices in
order to perform a consistent analysis.

We have defined four optimal observables ${\cal{O}}_{ttV}^\pm$ and
${\cal{O}}_{tbW}^\pm$ in eq.(\ref{observ}) which yield minimal
statistical errors in the determination of $C\!P$-violation
parameters. Therefore the statistical significance of the
non-standard signal is maximal. Adopting those observables, we have
presented in eq.(\ref{NSD2}), tables 2, and 3 the minimal values of
the $C\!P$-violation parameters which can be observed, as a function
of the luminosity ($L$) of the NLC and the achieved tagging
efficiency ($\epsilon_{tt}$). For $L=40\ {\rm fb}^{-1}$ and
$\epsilon_{tt}=15 \%$, one will be able to measure the
$C\!P$-violating parameters in the $t\bar{t}$ production and $t$
decay, i.e. $\xi$ and $\re{(f_2^R-\bar{f}_2^L)}$ respectively, at the
$3\sigma$ level if they are larger than 0.44 and 0.51, respectively.
It should be emphasized that {\it an observation of a non-zero signal
would be a strong evidence of physics beyond the SM.}

\vspace*{0.6cm}
\centerline{ACKNOWLEDGMENTS}

\vspace*{0.3cm}
We are grateful to K. Fujii for a kind correspondence on the tagging
efficiency $\epsilon_{tt}$, and to T. Hasuike for a valuable comment
on the original manuscript. This work is supported in part by the
Committee for Scientific Research (Poland) under grant 2\ P03B 180
09, by Maria Sk\l odowska-Curie Joint Fund II (Poland-USA) under
grant MEN/NSF-96-252, and by the Grant-in-Aid for Scientific Research
No.06640401 from the Ministry of Education, Science, Sports and
Culture (Japan).

\vspace*{0.6cm}
\noindent
{\bf Appendix}

The functions $f(x)$, $g(x)$, $\delta\! f(x)$
and $\delta g(x)$ are defined as follows:
\begin{eqnarray*}
&&f(x)=C_1 \Bigl\{\; r(r-2)+2x{{1+\beta}\over{1-\beta}}
-x^2 \Bigl({{1+\beta}\over{1-\beta}}\Bigr)^2\; \Bigr\}, \\
&&\hspace*{6.5cm}({\rm for\ the\ interval}\ I_1,\ I_4) \\
&&\phantom{f(x)}=C_1 \, (1-r)^2,
  \hspace*{3.02cm}({\rm for\ the\ interval}\ I_2) \\
&&\phantom{f(x)}=C_1 \, (1-x)^2,
  \hspace*{2.98cm}({\rm for\ the\ interval}\ I_3,\ I_6) \\
&&\phantom{f(x)}=C_1 \, x\Bigl\{\; x +{{4\beta}\over{1-\beta}}
-x \Bigl({{1+\beta}\over{1-\beta}}\Bigr)^2\; \Bigr\}, \\
&&\hspace*{6.5cm}({\rm for\ the\ interval}\ I_5)
\end{eqnarray*}

\begin{eqnarray*}
&&g(x)=C_2 \Bigl[\; -rx +x^2 {{1+\beta}\over{1-\beta}}
-x\ln {{x(1+\beta)}\over{r(1-\beta)}} \\
&&\phantom{g(x)}\ \ \ \ \ \ \ \
+{1\over{2(1+\beta)}}\Bigl\{\; r(r-2)+2x{{1+\beta}\over{1-\beta}}
-x^2 \Bigl({{1+\beta}\over{1-\beta}}\Bigr)^2\; \Bigr\}\; \Bigr], \\
&&\hspace*{6.5cm}({\rm for\ the\ interval}\ I_1,\ I_4) \\
&&\phantom{g(x)}
=C_2 \Bigl\{\; (1-r+\ln r)x +{1\over{2(1+\beta)}}(1-r)^2 \;\Bigr\},\\
&&\hspace*{6.5cm}({\rm for\ the\ interval}\ I_2) \\
&&\phantom{g(x)}
=C_2 \Bigl\{\; (1-x+\ln x)x +{1\over{2(1+\beta)}}(1-x)^2 \;\Bigr\},\\
&&\hspace*{6.5cm}({\rm for\ the\ interval}\ I_3,\ I_6) \\
&&\phantom{g(x)}
=C_2 \,x\Bigl[\; {{2\beta x}\over{1-\beta}}
-\ln{{1+\beta}\over{1-\beta}} \\
&&\phantom{g(x)}\ \ \ \ \ \ \ \
+{1\over{2(1+\beta)}}\Bigl\{\; x+{{4\beta}\over{1-\beta}}
-x \Bigl({{1+\beta}\over{1-\beta}}\Bigr)^2\; \Bigr\}\; \Bigr], \\
&&\hspace*{6.5cm}({\rm for\ the\ interval}\ I_5)
\end{eqnarray*}
where
$$
C_1\equiv {3\over{2W}}{{1+\beta}\over\beta},\ \ \ \ \ \ \ \
C_2\equiv {3\over{W}}{{(1+\beta)^2}\over\beta},
$$
and $\beta$, $r$ and $W$ are defined in the text ($\beta$ is given
after eq.(\ref{def-x}), and $r$ and $W$ are after
eq.(\ref{t-decay})). The intervals $I_i$ ($i=1\sim 6$) of $x$ are
given by
\begin{eqnarray*}
&&I_1:\ \ r(1-\beta)/(1+\beta) \leq x \leq (1-\beta)/(1+\beta), \\
&&I_2:\ \ (1-\beta)/(1+\beta)  \leq x \leq r,                   \\
&&I_3:\ \ r                    \leq x \leq 1, \\
&&\ \ \ \ \ \ \ \ \ \ \ \ \ \
        (I_{1,2,3}\ {\rm are \ for}\ r\geq (1-\beta)/(1+\beta)) \\
&&I_4:\ \ r(1-\beta)/(1+\beta) \leq x \leq r,                   \\
&&I_5:\ \ r                    \leq x \leq (1-\beta)/(1+\beta),\\
&&I_6:\ \ (1-\beta)/(1+\beta)  \leq x \leq 1.                   \\
&&\ \ \ \ \ \ \ \ \ \ \ \ \ \
        (I_{4,5,6}\ {\rm are \ for}\ r\leq (1-\beta)/(1+\beta))
\end{eqnarray*}

\begin{eqnarray*}
&&\delta\! f(x)
=C_3 \Bigl\{\; {1\over 2}r(r+8)-2x(r+2){{1+\beta}\over{1-\beta}}
+{3\over 2}x^2 \Bigl({{1+\beta}\over{1-\beta}}\Bigr)^2 \\
&&\phantom{\delta\! f(x)}\ \ \ \ \ \ \ \
+(1+2r)\ln{{x(1+\beta)}\over{r(1-\beta)}}\; \Bigr\}, \\
&&\hspace*{6.5cm}({\rm for\ the\ interval}\ I_1,\ I_4) \\
&&\phantom{\delta\! f(x)}
=C_3 \Bigl\{\; {1\over 2}(r-1)(r+5)-(1+2r)\ln r\; \Bigr\}, \\
&&\hspace*{6.5cm}({\rm for\ the\ interval}\ I_2) \\
&&\phantom{\delta\! f(x)}
=C_3 \Bigl\{\; {1\over 2}(x-1)(5+4r-3x)-(1+2r)\ln x\; \Bigr\}, \\
&&\hspace*{6.5cm}({\rm for\ the\ interval}\ I_3,\ I_6) \\
&&\phantom{\delta\! f(x)}
=C_3 \Bigl\{\; (1+2r)\ln{{1+\beta}\over{1-\beta}}
-{{4\beta x}\over{1-\beta}}(r+2)
+{{6\beta}\over{(1-\beta)^2}}x^2 \; \Bigr\}, \\
&&\hspace*{6.5cm}({\rm for\ the\ interval}\ I_5)
\end{eqnarray*}
\begin{eqnarray*}
&&\delta g(x)
=C_3 \Bigl[\; 1-\beta +2(3-\beta)r+{1\over 2}r^2
-{3\over 2}(1-2\beta)\Bigl({{1+\beta}\over{1-\beta}}\Bigr)^2 x^2 \\
&&\phantom{\delta\! f(x)}\ \ \ \ \ \ \ \
+(1+\beta)x\Bigl\{ {1\over r}(r-1)(3r+1)-{{2(r+2)}\over{1-\beta}}
\Bigr\} \\
&&\phantom{\delta\! f(x)}\ \ \ \ \ \ \ \
+\{ 1+2r+2(1+\beta)(r+2)x \}
\ln{{x(1+\beta)}\over{r(1-\beta)}}\; \Bigr], \\
&&\hspace*{6.5cm}({\rm for\ the\ interval}\ I_1,\ I_4) \\
&&\phantom{\delta\! f(x)}
=C_3 \Bigl[\; {1\over 2}(r-1)(r+5)-(1+2r)\ln r \\
&&\phantom{\delta\! f(x)}\ \ \ \ \ \ \ \
+(1+\beta)x\Bigl\{ {1\over r}(r-1)(5r+1)-2(r+2)\ln r \Bigr\}\;
\Bigr], \\
&&\hspace*{6.5cm}({\rm for\ the\ interval}\ I_2) \\
&&\phantom{\delta\! f(x)}
=C_3 \Bigl[\; -{7\over 2}-4r-\beta (2r+1)+2x\{1-\beta+r(2+\beta)\} \\
&&\phantom{\delta\! f(x)}\ \ \ \ \ \ \ \
+{3\over 2}(1+2\beta)x^2-\{2r+1+2(1+\beta)(r+2)x \}\ln x\; \Bigr], \\
&&\hspace*{6.5cm}({\rm for\ the\ interval}\ I_3,\ I_6) \\
&&\phantom{\delta\! f(x)}
=C_3 \Bigl[\; -(1+2r)\Bigl(2\beta-\ln{{1+\beta}\over{1-\beta}}\Bigr)
+{{6\beta^3}\over{(1-\beta)^2}}x^2 \\
&&\phantom{\delta\! f(x)}\ \ \ \ \ \ \ \
-2(r+2)x\Bigl\{ {{2\beta}\over{1-\beta}}
-(1+\beta)\ln{{1+\beta}\over{1-\beta}}\Bigr\} \; \Bigr], \\
&&\hspace*{6.5cm}({\rm for\ the\ interval}\ I_5)
\end{eqnarray*}
where
$$
C_3 \equiv {6\over W}{{1+\beta}\over\beta}{\sqrt{r}\over{1+2r}}.
$$

\vspace*{0.6cm}

\end{document}